\documentclass[prb,aps,twocolumn]{revtex4}

\usepackage{bm}
\usepackage{dcolumn}
\usepackage{graphicx}
\usepackage{amsmath}
\usepackage{color}

\begin{document}

%\preprint{APS/123-QED}
\title{Bond Disorder in Even-Leg Heisenberg Ladders}

\author{Kien Trinh}
\email{ktrinh@usc.edu}
\affiliation{Department of Physics and Astronomy, University of Southern California, Los Angeles, CA 90089, USA}
\author{Stephan Haas}
\affiliation{Department of Physics and Astronomy, University of Southern California, Los Angeles, CA 90089, USA}
%\author{Rong Yu}
%\affiliation{Department of Physics \& Astronomy, Rice University, Houston, TX 77005, USA}
%\author{Tommaso Roscilde}
%\affiliation{Laboratoire de Physique, CNRS UMR 5672, Ecole Normale Sup\'erieure de Lyon, Universit\'e de Lyon, 46 All\'ee d'Italie, Lyon, F-69364, France}

\date{\today}

\begin{abstract}
Random bond disorder in antiferromagnetic spin-1/2 Heisenberg ladders is investigated using the Stochastic Series Expansion Quantum Monte Carlo method. We find that the effects of individual bond impurities vary strongly, depending on their position on legs and rungs. We initially focus on how the distribution of local bond energies depends on the impurity concentration. Then we study how the phase diagram of even-leg ladders is affected by random bond doping. We observe Bose glass phases in two regimes ($h' < h \lesssim h_{c1}$ and $h'' < h <h_{c2}$) and a Bose-Einstein Condensate in-between. Their presence are discussed in the relation to the local bond energies. 
\end{abstract}
\maketitle

%\section{Introduction}
{\bf I. Introduction:} \ \ 
The effects of doping on spin-gap compounds has attracted considerable interest; following the discovery of new families of materials whose low-energy magnetic properties can be described by antiferromagnetic (AF) Heisenberg ladder models\cite{DagottoR96}. It is known that AF spin-$1/2$ Heisenberg ladders with an even number of legs are characterized by a singlet ground state. This symmetry of the ground state breaks down beyond a sufficiently high applied magnetic field or beyond a critical concentration of random dopants. In the former case, Bose-Einstein condensation (BEC) of spin triplets occurs in a magnetic field window, $h_{c1} < h < h_{c2}$, i.e. in an intermediate field regime which separates the spin-gap phase ($h<h_{c1}$) and a fully polarized phase ($h>h_{c2}$).\citep{Jaime04,Manakaetal08,Thielemann09}  In the latter case, one distinguishes two types of randomness: site and bond disorder. Site disorder induces local moments, which can give rise to long range order via an order-by-disorder mechanism. \citep{Shender91,Sigrist96,Yuetal08} In contrast, the presence of bond disorder can destroy magnetic long range order through quantum localization, leading to Bose glass formation\citep{Fisher89}, which has recently been observed in spin gap ladder compounds\citep{Manakaetal02, Trinh12}. Many interesting phenomena are induced by the simultaneous presence of both magnetic field and randomness.\cite{Roscilde05,Oosawa02} For example, in doped spin gap ladders one observes a very rich phase diagram, with a sequence of magnetic field controlled phases, including superfluidity, BEC, Bose glass and full polarization.\citep{Manakaetal08,Oosawa02,Yuetal10} Specifically, recent experiments on the doped compound IPA-Cu(Cl$_{0.95}$Br$_{0.05}$)$_3$ imply the existence of another Bose glass at low applied magnetic fields, before the appearance of BEC. While some of these results have been modeled quantitatively using a bond-disordered Heisenberg Hamiltonian \citep{Nohadani05} it still remains desirable to better understand the effects of random doping on local observables. 

In this paper, we study how bond disorder affects the quantum phase diagram of even-leg antiferromagnetic Heisenberg ladders in the presence of an applied magnetic field. Bond disorder occurs when dopant ions replace the ions which act as bridges between the magnetic ions. For instance, bond disorder can be introduced in IPA-$\mathrm{CuCl_3}$ by a partial substitution of non-magnetic $\mathrm{Br^-}$ for the likewise non-magnetic $\mathrm{Cl^-}$, affecting the bond angles in the $\mathrm{Cu}$-halogen-halogen-$\mathrm{Cu}$ superexchange pathways \cite{Manakaetal08, Hongetal10, Manakaetal09}. The enhanced strength of magnetic interactions on the affected bonds leads to a Bose glass when a sufficiently strong magnetic field is applied. Making use of Quantum Monte Carlo simulations we can address the local bond energies, uniform and staggered magnetization at ultralow temperatures. As discussed below, we find that with increasing magnetic field, BEC and Bose glass phases appear preceding the critical magnetic fields ($h_{c1}$ and $h_{c2}$) of the pure system. We study the influence of bond impurities on the bond energies $E_b$ of its neighbors. This allows us to characterize the effects of bond impurities on the condensation and localization of triplons, which occur in BEC and Bose glass phase respectively. In this context, a "superfluid" phase corresponds to the delocalization of bosons, and it is characterized by a non-zero magnetization $m_s^{\bot}$ perpendicular to the magnetic field. The Bose glass phase occurs due to the localization of triplons. It is characterized by a finite slope of the uniform magnetization $m_u$ and a vanishing order parameter $m_s^{\bot}$. 

%\section{Model}
{\bf II. Model:} \ \ 
We examine the bond-disordered antiferromagnetic Heisenberg model on 2- and 4-leg ladders described by the Hamiltonian
\begin{equation}\label{hamil}
H=\sum_{i,j}J_{ij}{\bf S}_{i} {\bf S}_{j} + h\sum_i S^z_{i},
\end{equation}
where ${\bf S}_{i}$ is the spin-1/2 operator at site $i$, $h$ is the applied magnetic field, and  $J_{ij}$ denotes the nearest-neighbor coupling between the spins on sites $i$ and $j$, taking the values $J=1$ with probability $1-p$ and $J'$ with probability $p$. Periodic boundary conditions are used along the leg direction. The simulations are performed using the Stochastic Series Expansion (SSE) Quantum Monte Carlo (QMC) method based on the directed loop algorithm \cite{SyljuasenS02}. In order to access the regime of very low temperatures, we use a $\beta-$doubling scheme \cite{Sandvik02}. This way, maximum inverse temperatures up to $\beta = 2048$ are obtained.

The staggered magnetization $m_s^{\bot}$ is calculated from the staggered structure factor,
\begin{equation}
S_s^{\bot}=\frac{1}{N}\sum_{\langle i,j \rangle}(-1)^{i+j}\langle S_i^x S^x_j \rangle, 
\end{equation}
using $m_s^\bot=\sqrt{S_s^{\bot}/N}$. And the local bond energies are measured by $E_{b} = J_{ij}\langle {\bf S}_i {\bf S}_j\rangle$. The staggered $m_{s}^{\bot}$ and uniform $m_u$ magnetizations are averaged over at least 700 impurity realizations. An ultra-low temperature $T=1/2048$ was chosen in all simulations so that the relevant thermodynamic observables reflect true zero-temperature behavior. The lengths of the 2-leg ladders used are $L_x=104, 128, 160, 200, 256, 320$, and $400$. And the lengths of the 4-leg ladders used are $L_x=96, 128, 160, 192, 256, 320$, and $400$. Their thermodynamic limit, i.e. infinite length, is extrapolated via finite-size scaling.

{\bf III. Bond-doped Heisenberg ladders:} \ \ Pure antiferromagnetic even-leg ladders display a resonant valence bond (RVB) ground state \cite{DagottoR96} with a finite gap $\Delta$ to triplet excitations, where the $n$ spins on the same rung preferentially form local singlets. The value of the spin gap decreases as the number of legs increases.\cite{White94} An interesting question that arises concerns how the magnitudes of the bond energies relate to the observed spin gap.

\begin{figure}[t]
%\centering
\includegraphics[scale=0.4,angle=-90]{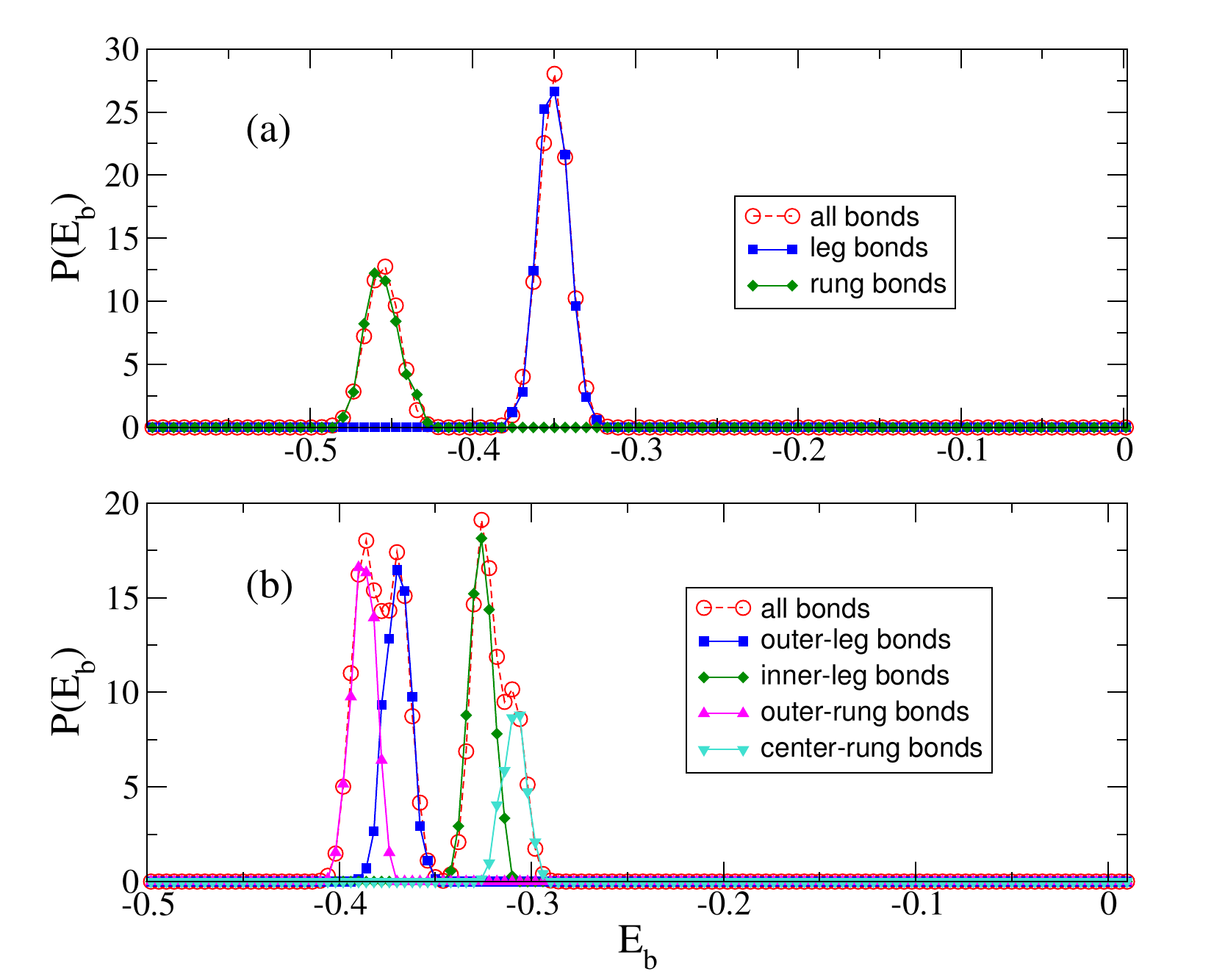}
\caption{\label{e_local}(color online) (a) Distribution of local bond energies in pure  (a) 2-leg and (b) 4-leg AF spin-1/2 Heisenberg ladders. For each case, there are different contributions due to rung bonds and leg bonds. Simulations are performed at temperature $T=J/2048$, and lattice length $L_x=256$.}
\end{figure}

Fig.~\ref{e_local} shows the local bond energy distribution in pure 2- and 4-leg ladders with $J_{ij} = J$.  The simulations are performed at a very low-temperature $T=J/2048$ on ladders with linear size $L_x=256$, measuring the local bond energies. In pure 2-leg ladders, the distribution function of these local bond energies shows two peaks \cite{explain1}, corresponding to two the types of bonds. The rung-bond energies give rise to the lower-energy peak at $E_b \approx -0.454J$, whereas the leg bonds cause the higher-energy peak at $E_b \approx -0.350J$. This observation is consistent with the fact that pure 2-leg ladders display a rung-singlet-dominated ground state. The local bond energies of pure 4-leg ladders fall into four categories,  corresponding to four types of bonds, located on outer rungs, outer legs, inner legs, and center rungs respectively. They are arranged in ascending order of local bond energies, given by $E_b \approx -0.388J,-0.370J,-0.326J$, and $-0.309J$. It is noted that the center-rung bonds have the highest energies. In other words, the four rung spins preferentially form effective pairs of weakly coupled singlets on each rung.

\begin{figure}[t]
%\centering
\includegraphics[scale=0.25]{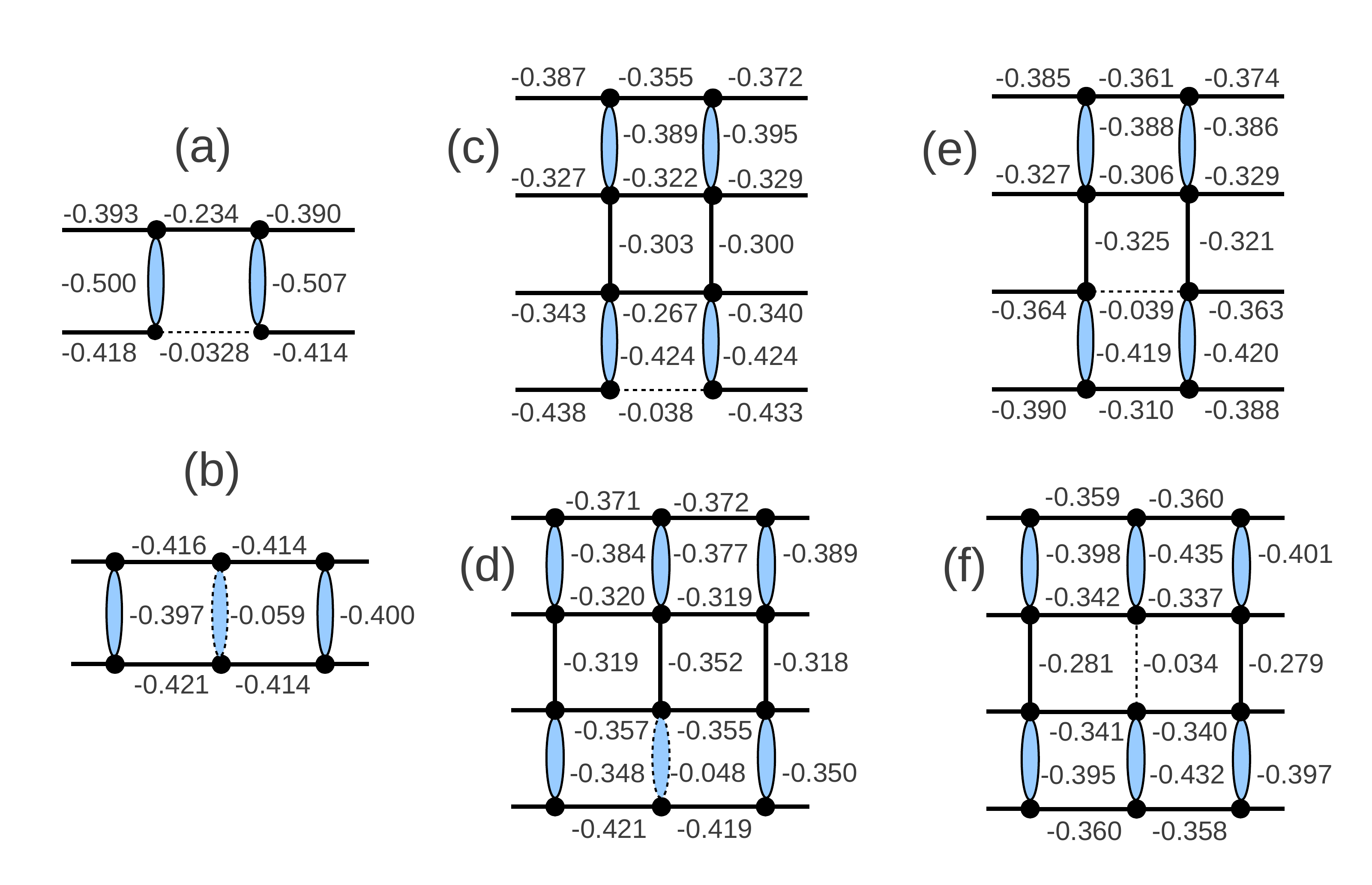}
\caption{\label{e_local_diagram}(color online) Effects of local bond impurities, denoted by dashed lines, on the neighboring bond energies in 2- and 4-leg ladders. Simulations are performed at temperature $T=J/2048$, lattice length $L_x=256$, with the coupling of the bond impurity $J'=J/5$.}
\end{figure}

What happens when the ladders are bond-doped? For sufficiently low bond dilution concentrations, the average distance between the individual bond impurities is large. To understand this regime, it is useful to study the local effects of isolated impurity bonds. Fig.~\ref{e_local_diagram} shows the effect of an isolated  impurity on its neighbor bonds. Here, one bond is replaced by a weaker coupling $J'=J/5$, and its position is varied. In the case of 2-leg ladders, the impurity bond can be located on either a rung or on a leg. In Fig.~\ref{e_local_diagram}(a), it is observed that a leg-bond impurity significantly reduces the magnitude of the bond energy on the opposite leg bond. Additionally, it strongly enhances the bond energies on its two neighbor rungs, indicating that this type of bond impurity strengthens the formation of rung singlets in its immediate neighborhood. In contrast, when the bond impurity is placed on a rung, it  reduces the magnitude of the bond energies of the two closest rungs, see Fig.\ref{e_local_diagram}(b), but increases the magnitude of the bond energies on the neighboring legs. 

For the case of 4-leg ladders, the preferential formation of pairs of effectively weakly coupled singlets along the 4-site rungs has significant consequences when bond doping is introduced. As discussed above, an approximate effective description of the ground state in a pure 4-leg ladder would be a product state of two weakly coupled 2-leg ladders. When a  bond impurity is introduced on a leg, it thus causes an effect similar to the 2-leg ladder case, see Fig.~\ref{e_local_diagram}(c) and (e), i.e. the opposite leg bond energy on the 2-leg subsystem on which the impurity is placed is reduced while bond energies on neighboring rungs are enhanced. The other 2-leg subsystem is barely affected by the leg impurity.  In contrast, if the impurity is introduced on an outer rung, it causes a reduction of the rung neighbor bond energies on the same 2-leg ladder subsystem, see Fig.~\ref{e_local_diagram}(d). It also enhances the bond energies on the neighboring leg bonds, which again is similar to the effect in Fig.\ref{e_local_diagram}(b). Last but not least, when the impurity is placed on a center rung, it enhances the bond energies on the two neighboring  rungs. However, its influence compared with the previous case is much less drastic, see Fig.~\ref{e_local_diagram}(f).

\begin{figure}[t]
%\centering
\includegraphics[scale=0.4,angle=-90]{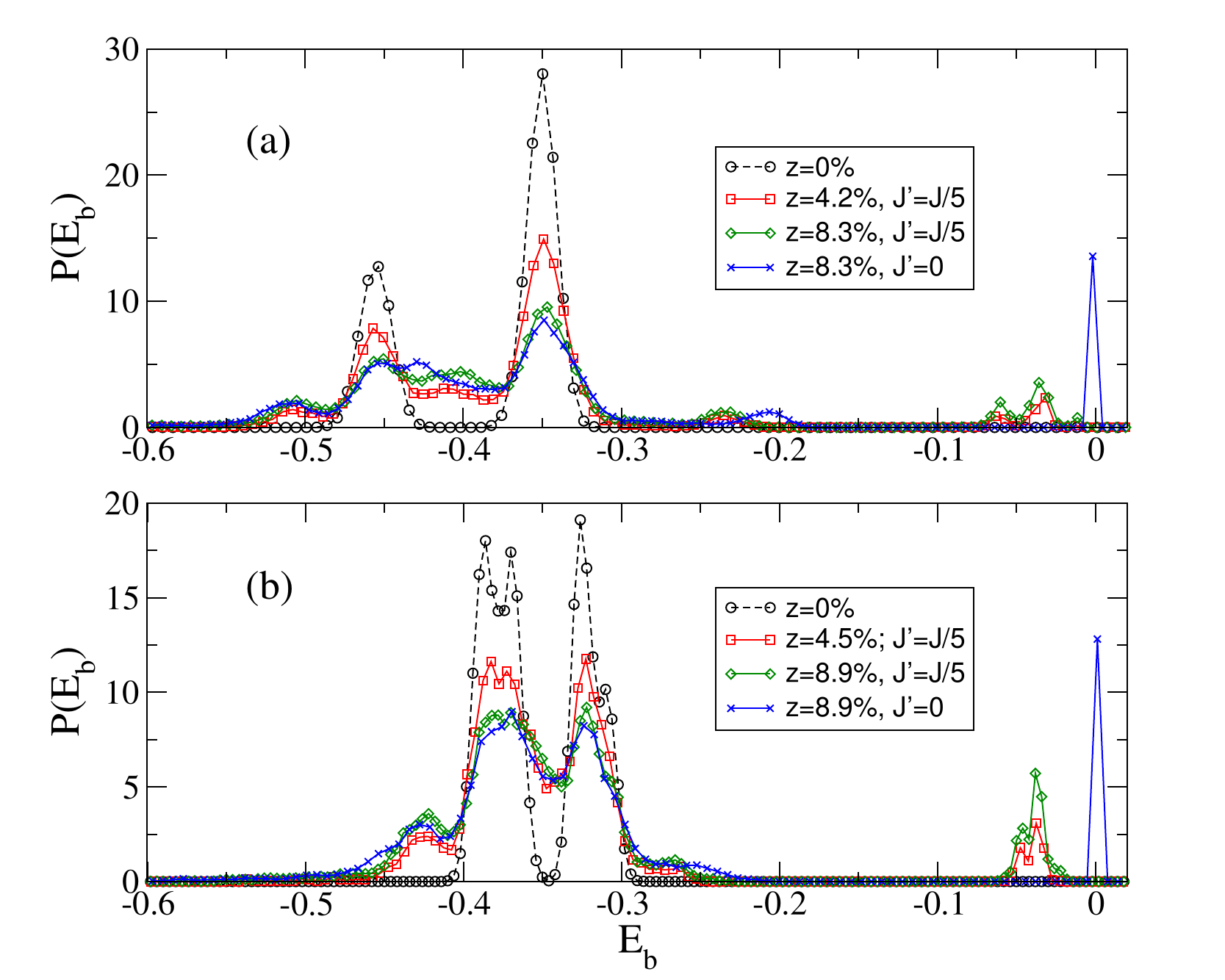}
\caption{\label{e_local_doped}(color online) (a) Distribution of local bond energies in bond-disordered (a) 2-leg and (b) 4-leg  ladders. Simulations are performed at temperature $T=J/2048$, and lattice length $L_x=256$.}
\end{figure}

In randomly bond doped ladders, the distribution of local bond energies displays characteristics stemming from the effects discussed above for isolated impurity bonds. Fig.~\ref{e_local_doped} shows the distribution of local bond energies of doped (a) 2-leg and (b) 4-leg ladders. Here, we show three typical examples with $(z=4.2\%, J'=J/5)$, $(z=8.3\%, J'=J/5)$, and $(z=8.3\%, J'=0)$. 

For the 2-leg ladder, it is observed that, besides the main peaks of the pure case, there are three additional features. For instance, let us consider the case $(z=8.3\%, J'=J/5)$. Here, the additional peaks appear at energies  $\approx -0.51J$, $-0.22J$, and $-0.05J$. These features originate from affected bonds in the vicinity of the bond impurities. The lowest-energy peak is the contribution of singlets next to leg-bond impurities, the second lowest-energy peak comes from leg bonds opposite to the impurity, and the third peak stems from the bond-impurities themselves on the rung and leg bonds. There is actually another feature in-between the two main peaks, which stems from rung singlets next to the rung-bond impurities. 

In 4-leg ladders with $(z=8.9\%, J'=J/5)$, the feature at $-0.42J$ stems from the rung bonds shown in Figs.~\ref{e_local_diagram}(c), (e), and (f) and outer-leg bonds in Fig.~\ref{e_local_diagram}(d). The peak at $E_b \approx -0.27J$ is corresponds to center leg couplings in Fig.~\ref{e_local_diagram}(c). The slightly lower energy peak at $E_b \approx -0.28J$ is a contribution of center rung bonds next to rung-bond impurities, see Fig.~\ref{e_local_diagram}(f). The peak at $-0.03J$ is the contribution of bond-impurities. Clearly, the local bond energy distribution is much richer compared with the 2-leg ladder case, but the individual features can still be explained using a local description. 

\begin{figure}[t]
%\centering
\includegraphics[scale=0.4,angle=-90]{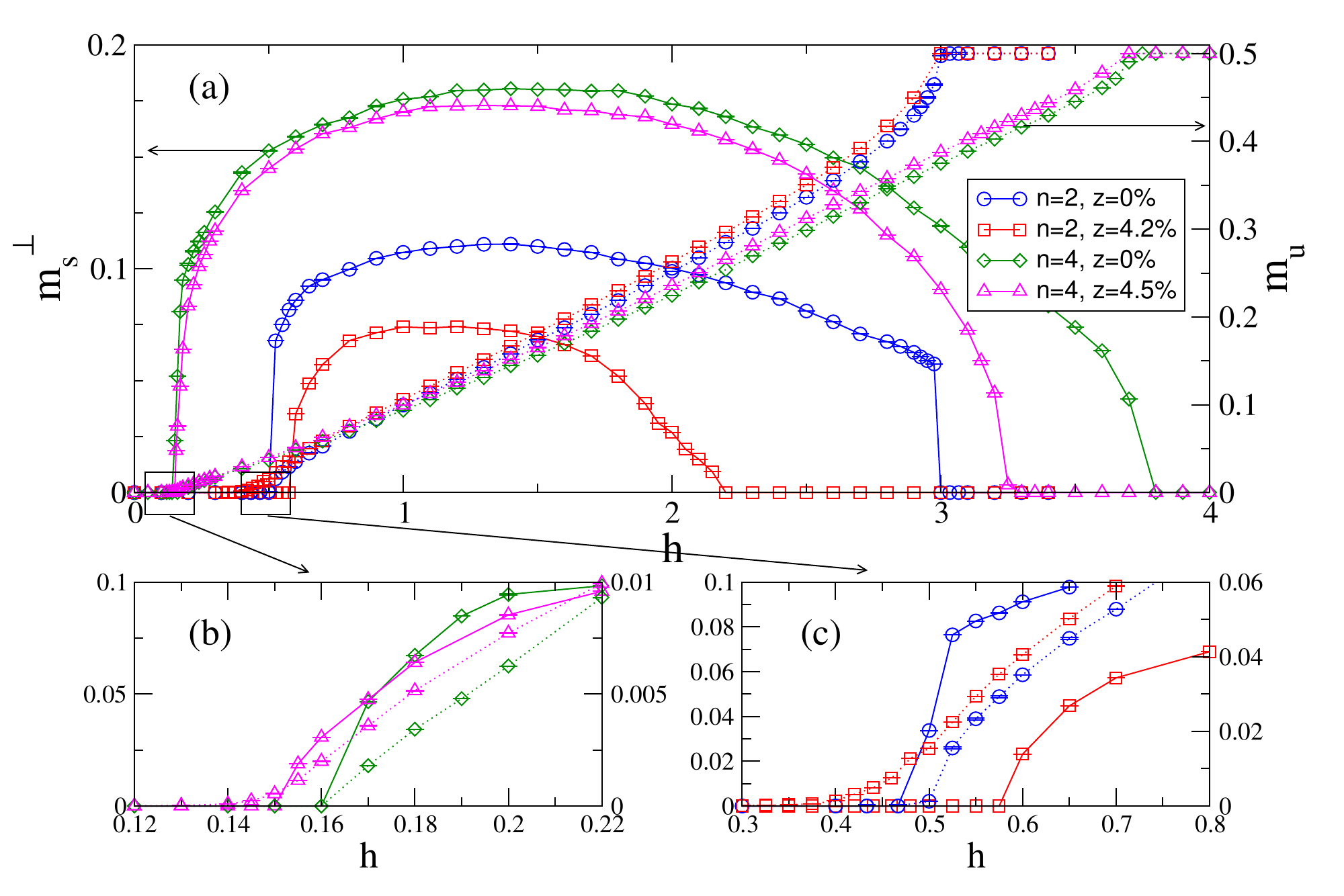}
\caption{\label{3j3z}(color online) Staggered and uniform magnetization of pure and bond-disordered even-leg ladders. Bond couplings are randomly replaced by couplings $J'=J/5$. The values of the magnetization are averaged over at least 700 realizations for each lattice size. The uniform magnetization is measured at lattice length $L_x=256$. The thermodynamic limit of the staggered magnetization, i.e. infinite length, is extrapolated via finite-size scaling using $L_x=104, 128, 160, 200, 256, 320$, and $400$ for 2-leg ladders and $L_x=96, 128, 160, 192, 256, 320$, and $400$ for 4-leg ladders. The error bars fall within the symbol size.}
\end{figure}

\begin{figure}[t]
%\centering
\includegraphics[scale=0.3]{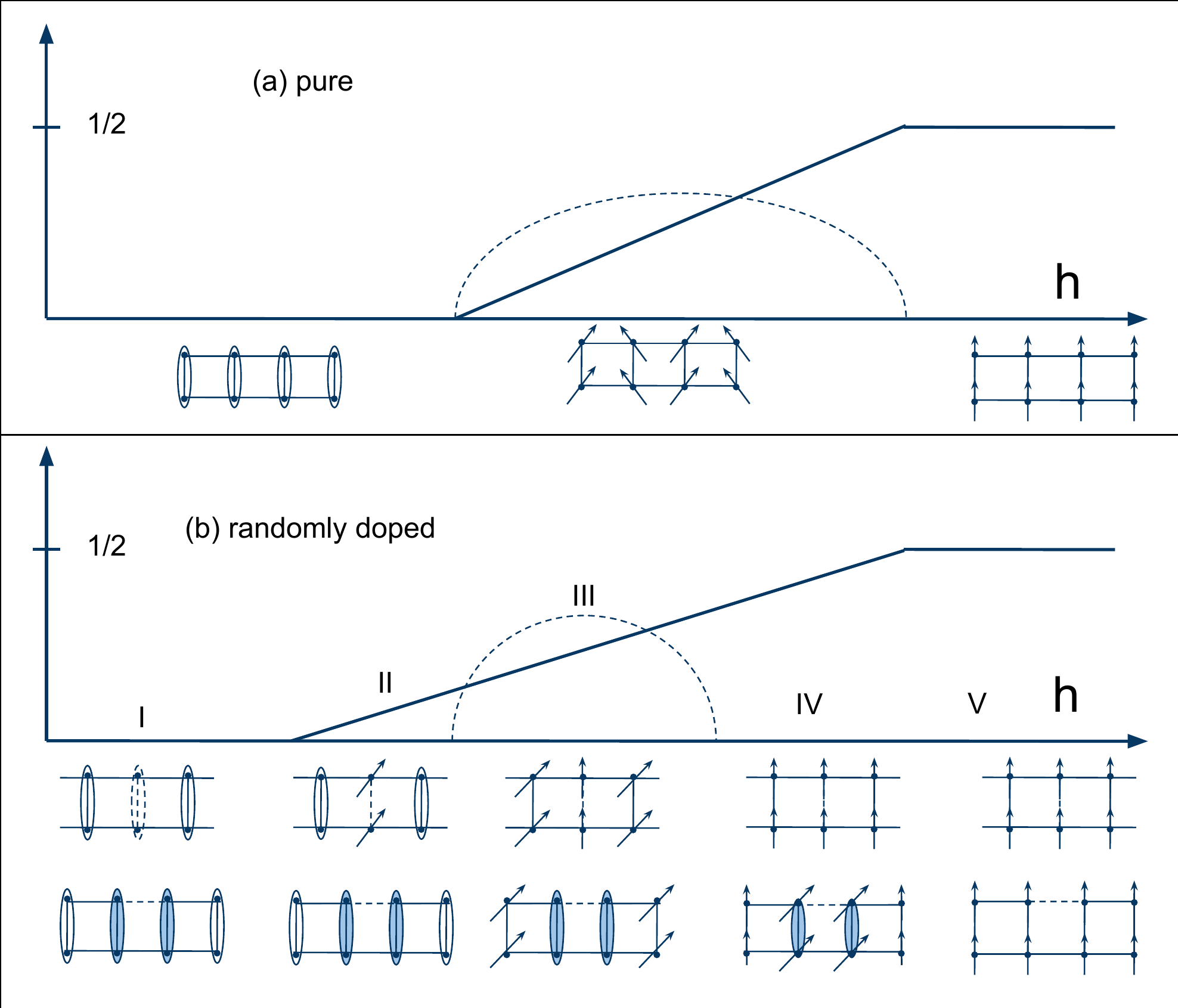}
\caption{\label{states}(color online) Schematic illustration of the response of the uniform and staggered magnetizations to an applied magnetic field. In the pure case (a), the system remains in an RVB spin fluid state up to a lower critical field $h_{c_1}$. Then it undergoes a BEC transition into the regime $h_{c_1} < h < h_{c2}$.  All spins are fully polarized beyond the saturation field $h_{c2}$. In the doped case (b), a field scan reveals the following sequence of phases:  I: RVB spin liquid; II: Bose glass phase; III: BEC; IV: Bose glass phase; V: fully polarized phase.}
\end{figure}

The strength of local bond energies has direct consequences for the phase diagram when the randomly bond doped ladders are exposed to an applied magnetic field. Fig.~\ref{3j3z} shows the quantum (low-temperature) phase diagram of pure and doped 2 and 4-leg ladders in the presence of a uniform magnetic field $h$. The uniform magnetization is found  not to depend significantly on lattice size. Therefore only $L_x=256$ is shown. The thermodynamic limit of the staggered magnetization is calculated by an extrapolation of the finite-size data $m_s^{\bot}(L)$ as a function of $L$. For instance, in the spin-gap, Bose glass and fully polarized phases a linear scaling in $1/L$ is used, whereas in the BEC phase the extrapolation requires the use of higher order polynomials.\citep{Nohadani05b} The presence of these phases will become clear later. In the pure case, for sufficiently small fields, both the uniform $m_u$ and staggered $m_s^\bot$ magnetization remain. They  become finite beyond a lower critical field, $h_{c1}=0.5$ for the 2-leg ladder and $h_{c1} = 0.16$ for the 4-leg ladder. Spins in pure 2-leg ladders are fully polarized at an upper critical magnetic field $h_{c2}=3$, and similarly, the upper critical field for 4-leg ladders is $h_{c2}=3.8$.\citep{Wessel01} Fig.~\ref{states}(a) provides a schematic picture to explain the observation of the phase diagram found from the QMC simulations. At zero field, quantum spin fluctuations in the pure 2-leg Heisenberg ladder destroy conventional magnetic order. The RVB ground state can be approximated by a product state of rung spins  forming singlets, separated from the lowest triplet excitation by a minimum excitation energy $\Delta$.\citep{DagottoR96} This is a non-magnetic spin liquid state. The spin gap is reduced as the magnetic field increases. It closes when the magnetic field reaches the lower critical value $h_{c1}$, at which the system experiences a quantum phase transition from the spin liquid phase to a BEC of magnons.\citep{Nohadani05} This phase is characterized by non-zero uniform and staggered magnetizations. Ultimately, at large applied magnetic fields all the spins are polarized along the $z$-direction, i.e. for $h>h_{c2}$. At this upper critical magnetic field, the system transitions from a BEC to a polarized paramagnet, and there is no longer coherence between spins. The staggered magnetization vanishes and the uniform magnetization saturates at $m_u=1/2$.

In the doped case, these phase diagrams are modified by the presence of bond disorder. The staggered magnetization of doped 2 and 4-leg ladders are shown for respective doping concentrations $z=4.2\%$ and $4.5\%$. The bond couplings $J$ are randomly replaced by $J'=J/5$. This particular choice is motivated by recent experiments,\citep{Manakaetal09} but the results shown here are generic. Interestingly, close to both critical fields $h'<h \lesssim h_{c1}$ and $h''<h<h_{c2}$ (Fig.~\ref{3j3z}(b) and (c)) we observe a finite slope of the uniform magnetization but a vanishing staggered magnetization. These are the characteristics of Bose glass phases. We observe the phase transition of these Bose glass phases to BEC at a magnetic field smaller than $h_{c1}$ for 2-leg ladders and greater than $h_{c1}$ for 4-leg ladders. Observations of disorder induced phases near the originally critical fields, recently reported in three-dimensional dimer systems, \citep{Nohadani05} suggest that BEC and Bose glass phases appear near the higher critical fields. Our finding differs in that for the bond-doped 2-leg ladder the Bose glass phases occur at both critical fields, depending on the magnitude of bond disorder. However the existence of Bose glass phases at both critical magnetic fields has recently been observed in the compound IPA-Cu(Cl$_{0.95}$Br$_{0.05}$)$_3$. \citep{Hongetal10} Below, we discuss mechanisms derived from our model study which offer explanations for the experimental observations. 

In Fig.~\ref{states}(b), we show the five phases which appear in the QMC data of {\em doped} 2-leg ladders. At sufficiently small fields the system stays in the spin liquid regime I. A finite magnetic field strength is required to overcome the lowest singlet-triplet gap of bond-impurity couplings. As discussed above, the impurity rung-bonds are weaker, and therefore singlets on these bonds break first, when an external magnetic field is applied to the system. Beyond this first critical field, the uniform magnetization becomes finite with increasing magnetic field while the staggered magnetization stays zero. This magnetic field is sufficiently strong  to gradually polarize spins on the impurity rung-bonds, but not strong enough to polarize the spins in the \emph{bulk}. During this phase, the field-induced triplons stay localized on the impurity-rung bonds. There is no coherence among them. The system therefore forms a Bose glass phase which is manifested in region II. At magnetic field, $h \approx h_{c1}$, singlets on the \emph{bulk rungs} break, triplons form on these rung-bonds and they interact with each other. We note that for any small amount of dilution impurities in a 2-leg ladder separate rung-bonds into clusters. Rung couplings on these clusters are generally greater than that in the pure case. Hence, magnetic fields higher than $h_{c1}$ are needed to polarize these rung-bonds, Fig.\ref{3j3z}(c). However, the above argument can not be applied to 4-leg ladders because at small dilution rung-bonds are not separated into clusters. Therefore a magnetic field smaller than $h_{c1}$ is sufficient to polarize these rung-bonds, Fig.\ref{3j3z}(b). The system undergoes a BEC phase transition, regime III. Recall that there are still basically un-renormalized singlets next to the leg-bond impurities untouched. Their bond energies are relatively large compared with the original bond energies, see Fig.~\ref{e_local_diagram}(a). Although the magnetic field has fully polarized the original singlets in this regime, it is not sufficiently strong to break these impurity induced higher-energy singlets. The system therefore undergoes another Bose glass phase transition into regime IV. Finally, above $h_{c2}$, all spins finally polarized, regime V. This sequence is observed in the QMC data shown in Fig.~\ref{3j3z}.

%\section{Conclusions}
{\bf IV. Conclusions:} \ \ 
To summarize, we have studied the effects of bond impurities in even-leg antiferromagnetic spin-1/2 Heisenberg ladders. We find that depending on the location of the bond impurities, bond energies on neighboring bonds  are either enhanced or reduced. In general, bond impurities enhance bond energies connected with them by the same spins, and reduce the opposite couplings. This effect is related to the so-called edge disorder effect. \citep{Zhangetal10} We identify various types of impurity induced bond energy shifts in 2- and 4-leg ladders. In light of these results, we demonstrate the emergence of BEC and Bose glass phase close to both critical fields of the pure system. These results can be used to explain recent observations of a disorder induced Bose glass phase in IPA-Cu(Cl$_{0.95}$Br$_{0.05}$)$_3$. \citep{Hongetal10}

\begin{acknowledgments}
We would like to thank Tommaso Roscilde and Rong Yu for valuable discussions. We also acknowledge financial support by the Department of Energy, grant number DE-FG03-01ER45908. The numerical computations were carried out on the University of Southern California high-performance supercomputer cluster.
\end{acknowledgments}

\end{document}